# Evidence for the Domain Wall Nature of Sketched LaAlO$_3$/SrTiO$_3$ Nanowires

Dawei Qiu[1], Mengke Ha[1], Qing Xiao[1], Zhiyuan Qin[1,2], Danqing Liu[1], Changjian Ma[1], Guanglei Cheng[1,2†] and Jiangfeng Du[1,2‡]

[1]*CAS Key Laboratory of Microscale Magnetic Resonance and School of Physical Sciences, University of Science and Technology of China, Hefei 230026, China*

[2]*Hefei National Laboratory, Hefei 230088, China*

[†]glcheng@ustc.edu.cn, [‡]djf@ustc.edu.cn

**Abstract:**

The rich electron correlations and highly coherent transport in reconfigurable devices sketched by a conductive atomic force microscope tip at the LaAlO$_3$/SrTiO$_3$ interface have enabled the oxide platform an ideal playground for studying correlated electrons and quantum technological applications. Why these one-dimensional devices possess enhanced properties over the two-dimensional interface, however, has remained elusive. Here we provide evidence that one-dimensional LaAlO$_3$/SrTiO$_3$ nanowires are intrinsically ferroelastic domain walls by nature through thermodynamic study. We have observed spreading resistance anomalies under thermo-stimulus and temperature cycles, with characteristic temperatures matching domain wall polarity. This information is crucial in understanding the novel phenomena including superconductivity and high mobility quantum transport.

**Main text:**

The exploration of emergent phenomena at correlated oxide interfaces has been a major thrust of oxide research over the past two decades[1]. The $LaAlO_3/SrTiO_3$ (LAO/STO) interface, as a prime example, hosts a series of tunable phases involving superconductivity[2], metal-insulator transition (MIT)[3], spin-orbit interaction[4,5], and magnetism[6]. These novel phases can be further programmed to a nanoscale using a sharp conductive atomic force microscope (cAFM) tip[7,8]. Through reconfigurable "writing" and "erasing" procedures at the insulating 3 unit-cell (uc) LAO/STO interface, devices like superconducting nanowires[9,10], single electron transistors (SketchSET)[11,12], ballistic electron waveguides[13,14], and Fabry-Perot electron resonators[15] can be readily created with an unprecedented resolution and display a plethora of surprisingly clean quantum mechanical effects and correlation driven phenomena. The combination of precise reconfigurability, rich electron correlations and tunable quantum effects makes the oxide interface an attracting platform for quantum applications including Majorana particle based topological quantum computing and solid-state quantum simulations[16].

The origin of the novel physical behaviors in these 1-dimensional (1D) cAFM sketched devices, however, remains largely elusive, especially when compared with the 2-dimensional (2D) LAO/STO interface. For examples, the mobility of sketched nanowires easily exceeds 10000 $cm^2$/Vs, nearly an order of enhancement compared to the typical value (~1500 $cm^2$/Vs) at the 2D LAO/STO interface[17]. Similarly, the mean free path at 1D sketched devices can reach tens of micrometers, enabling quantized conductance and Fabry-Perot resonance to be observed over a long range, in stark contrast to the sub-100 nm mean free path at 2D LAO/STO interface[13,15,18]. The superconductivity at 2D

LAO/STO interface is believed to be weak Bardeen-Cooper-Schrieffer (BCS) type, at least when the carrier density is relatively high (range in $10^{13}$ cm$^{-2}$). However, transport experiments of sketched nanowires reveal that the superconductivity is intrinsically quasi-1D. The critical current in the sketched nanowires seems independent of wire width; rather, it scales with the number of wires[19]. Additionally, transport experiments in nanowire-based superconducting SketchSET show characteristic tunneling behaviors of single electrons and pairs, revealing the existence of electron pairs at effective temperatures (1~10 K) far above the critical temperature in the clean limit of bulk STO (~0.3 K) and magnetic fields (1~11 T) far beyond the upper critical fields (~0.2 T) of 2D LAO/STO interface[12]. Subsequent experiments in electron waveguides have observed a Pascal electron liquid phase with electron bundling to pairs and trions below large critical magnetic fields[18]. These experiments point to a real-space pairing mechanism by which local attractive interaction mediates electron pairing and helps form the Pascal liquid.

What causes the qualitative differences between cAFM sketched nanowires and the 2D LAO/STO interface? The cAFM tip-induced metal-insulator transition is in principle, an electronic charge transfer process similar to the 2D interface[20]; however, additional structural distortion is clearly observed in the piezoresponse force microscopy (PFM) at room temperature[21], which reveals lattice elongation along $z$ direction over the cAFM sketched area. Such lattice distortion is reminiscent of the ferroelastic domain walls formed after the cubic to tetragonal structural transition at $T_{c1}$=105 K in STO[22,23]. Through extensive imaging and spectroscopic studies, the consensus on the domain wall physics in STO lies as follows: the domain wall is polar below $T_{c2}$=80 K, with polarity increasing rapidly below $T_{c3}$=40 K[24,25]; no domain wall pinning is observed at low temperatures, instead, the domain wall in STO is highly mobile and tunable by electric gating and stress[24,26,27]; in addition, the conductivity and superconducting critical currents are higher on the domain wall than the surrounding bulk, as revealed by scanning superconducting quantum interference device (SQUID) imaging[28] and anisotropic transport measurement[29]. It is also argued that the high mobility carriers at the 2D interface move preferentially along the 1D domain walls, dominating the Shubnikov-de Hass oscillations seen at the 2D interface[14]. Lastly, the ferroelastic waves are theoretically proposed to be responsible for the real-space attractive electron pairing and the 1D nature of superconductivity of sketched nanowires[30]. Hence, whether the cAFM sketched nanowires can seed formation and bear properties of ferroelastic domain walls is crucial to understand the novel underlying physics.

Here we investigate the thermodynamic responses of cAFM sketched nanowires and reveal characteristic signatures of ferroelastic domain wall nature. We structure the experiments to answer the following two questions. 1) Do the cAFM nanowires behave like the ferroelastic domain walls? 2) If they do, are the signatures coming intrinsically from the wires themselves or extrinsically from intersecting domain walls in bulk STO (although these domain walls are insulating)? As has been visualized by numerous local imaging techniques, the ferroelastic domains are ubiquitous at 2D LAO/STO interface below $T_{c1}$=105 K, and are mainly classified as *X, Y,* and *Z* domains orienting $\pm 45°$ or parallel to the crystallographic axes[26,28,31]. The *Z/X,Y* domain walls typically separate around 1~10 $\mu m$ and their polarity, as revealed by scanning SET microscopy[26], may cause potential fluctuations across the interface and scatter electrons. Hence they are the primary targets of study in this work. We set out to write the cAFM nanowire (device A) along a hexagonal path on insulating 3-uc LAO/STO samples as shown in Fig. 1(a); more details on sample growth and cAFM fabrication methods can be found in the **APPENDIX A**. This hexagonal geometry allows 5 sections of wires (3 $\mu m$ long, 10 nm wide) to have a good chance of interacting with the $Z/X,Y$ domain walls. PFM imaging at room temperature further reveals the traces of the wires, signifying $z$ lattice distortions which are expected to seed the domain wall formation at low temperatures (Fig. 1(b)).

Figure 1(c) shows the temperature dependence of the resistance values $R_1$, $R_2$, $R_3$ $R_4$ and $R_5$ of wire sections 1 to 5, which nominally show similar behaviors. The smoothness in the cool-down curve, especially at $T_{c1}=105$ K, suggests there is no first-order signature related to the domain walls, although random resistance jerks do show up in some devices at $T<105$ K (**APPENDIX B** Fig. 5).

Interestingly, the temperature dependence of resistance ratios $r_{mn}(T) = R_m(T)/R_n(T)$ between different sections of wires reveals correlations to the ferroelastic domain wall formation, where $m$ and $n$ denote section index. Resistance ratios remove common contributions of resistance from mechanisms including scattering by immobile defects, electron-phonon coupling and electron-electron interactions. The contribution from possible scattering from domain walls, however, will vary from different cool-downs since the ferroelastic domains redistribute by cycling the temperature above 105 K. Figures 2(a) and 2(b) show the typical temperature-dependent normalized ratios $\tilde{r}_{53}$ and $\tilde{r}_{12}$ in 17 temperature cycles from 160 K to 20 K, where $\tilde{r}_{mn}(T) = r_{mn}(T)/r_{mn}(160 \text{ K})$. More data from other combinations of wire sections can be found in the **APPENDIX C** (Fig. 6). As a general observation from all the 17 temperature cycles, $\tilde{r}_{mn}(T)$ faithfully repeats above a critical temperature $T_{c2}=80\sim90$ K while spreading out below $T_{c2}$ in different cool-downs.

The insets in Figs. 2(a) and 2(b) show this spread $\Delta_{mn}(T) = \tilde{r}_{mn}^{max}(T) - \tilde{r}_{mn}^{min}(T)$ between maximum and minimum $\tilde{r}_{mn}$ values, which clearly illustrate this trend. A more detailed look shows the spread increases more rapidly below $T_{c3}$=40 K after emerging at $T_{c2}$. These temperatures match well with the characteristic temperatures of domain wall polarity, which first appears at 80 K and strengthens below 40 K, suggesting a direct correlation between cAFM nanowires and domain walls. However, whether this correlation arises from the cAFM nanowires intrinsically or by intersecting domain walls extrinsically in STO bulk remains a question by far.

Further insights can be gained by analyzing the distribution of $\tilde{r}_{mn}(T)$. Since the length of cAFM nanowires and domain width (1~10 $\mu m$) are comparable, $\tilde{r}_{mn}$ should rather take discrete values if each intersection of cAFM nanowires and domain walls causes observable resistance change (see the inset in Fig. 2C for an illustration). Namely, if we denote $(i,j)$ as the number of times that a nanowire section $m$ ($n$) intersects by random domain walls, then $\tilde{r}_{mn}$ in different temperature cycles would take values on the most probable configurations (0,0), (0,1), (1,0) etc. As a result, one would expect the distribution function of $\tilde{r}_{mn}$ peaks at specific $\tilde{r}_{mn}$ values. Namely, the cool-down curves should cluster into several bunches if plotted together. Figure. 2(c) reflects the distribution function by binning $\tilde{r}_{mn}$(50 K) from 17 temperature cycles and counting the occurrence frequencies. For all the possible $\tilde{r}_{mn}$ ($\tilde{r}_{31}, \tilde{r}_{15}, \tilde{r}_{52}$ shown in Fig. 2c) values; however, the occurrence frequencies are mostly singly peaked rather than peaking at multiple $\tilde{r}_{mn}$ values. This suggests the domain wall characteristics of cAFM nanowires are less likely to arise from intersecting the micrometer-spaced Z/X,Y domain walls in STO.

We note that instead of using the resistance ratios, the resistance difference between different cool-down curves works equally well in analyzing the above anomalous behaviors (see **APPENDIX D** Fig. 7). In addition, the above analysis mainly applies to the micrometer-sized $Z/X,Y$ domains that are detectable by scanning SET. It is well-known that nanoscale domains also exist in STO, with an excellent example of needle domains[32]. However, in a similar experiment at a conducting 2D LAO/STO Hall bar, we do not find such nanoscale domains cause observable resistance spread (Fig. 10). This can be understood that although the domain walls redistribute in each temperature cycle, their impact on the resistance at the 2D interface averages out. It would follow the same argument if nanoscale domain walls merely intersect with the nanowire, so no resistance spread should be observed unless the lattice along the nanowire direction is heavily distorted. Namely, the nanowire itself is formed by domain walls.

In the next, we explore further to answer whether the cAFM nanowire is a single continuous domain wall or comprises of many sections of domain walls in series. This is very important to the quantum transport in cAFM nanodevices. We can gain insights from the fabrication process of ballistic electron waveguides, in which an extra writing step with small voltages (~1 V) along the same path written at a much higher voltage (~10 V) is found always needed to ensure the emergence of conductance quantization[13,14,18]. This additional writing step is not able to bring more carriers to the wire, but is postulated to effectively remove scattering centers in the wire. Additionally, random resistance switching is often observed below 80 K in cAFM wires, as shown in **APPENDIX B** Fig. 5. These two empirical experimental facts can be understood if the cAFM nanowire is formed by a chain of individual domain walls. Under this scenario, the extra small-voltage writing should presumably merge all domain walls, and random domain wall switching causes significant resistance jerking.

Generally, a collective ferroelastic domain system is quite complex and shows nonlinear dynamics like quenching, jerking, and exponential relaxation in the presence of external stress, electric and thermal stimuli[24,29,33-36]. Pesquera et al. reported glassy behaviors of ferroelastic domain walls in bulk STO through resonant piezoelectric spectroscopy (RPS) and resonant ultrasound spectroscopy (RUS)[36]. The glassiness is manifested with exponential decay of a function related to resonance frequencies in RPS and RUS, corresponding to pinning and depinning of domain walls upon thermal activation below ~50 K[24]. Ojha *et al.* studied the electric stimulus by back gating a low mobility conducting $\gamma$-$Al_2O_3$/STO interface and revealed an exponential relaxation of sheet resistance owing to trapping and de-trapping of oxygen vacancies on the ferroelastic domain walls[37]. Here, we take an approach by applying a thermal stimulus to cAFM nanowires at different temperatures and look at the characteristic thermodynamic behaviors. The device (device B) is a straight wire written at 15 V (**APPENDIX E** Fig. 8) which displays an up-turn in resistance below $T$=35 K. At each temperature (from 20 K to 120 K in step of 10 K), the heater on the sample holder in the cryostat is turned off first to initiate a slight trend of device cooling, then it is turned back on with full power (~40 W) to reverse the temperature ramping direction. Surprisingly, although the temperature responds in no delay with heat power, the resistance of the nanowire shows dramatic opposite changes compared to the trend on the cool-down curve (Fig. 3). For example, at $T$=40 K in Fig. 3(c), the wire resistance is expected to increase from 37 $k\Omega$ in a rough trend marked by the blue dashed line (check Fig. 9 for extraction) after the heater is turned back on; however, it drops rapidly to 27 $k\Omega$ (27% change) first then slowly recovers to the expected values. Furthermore, this effect is more pronounced at lower temperatures and disappears at *T*>80 K.

Figure 4 shows the time evolution of the anomalous resistance drop $\Delta_r = \frac{R_a - R_n}{R_n}$ at each temperature stage, where $R_n$ and $R_a$ denote the resistance expected from the cool-down curve and actual resistance, respectively. Note that the time axis is normalized by the total time during which the resistance anomaly lasts at each temperature. Clearly, $\Delta_r$ shows the largest drop ($\Delta_m$) after the heater is turned on, then decays slowly in a fashion that can be well fitted by an exponential function $\Delta_r = \Delta_m e^{-t/\tau}$ (Fig. 4(a) inset) below *T*=50 K, where $\tau$ is the relaxation constant. This agrees with RUS and RPS studies from Pesquera *et al.* and Scott *et al.*, that ferroelastic domain walls soften with enhanced domain wall mobility and polarity in this temperature range[24,36]. Figure 4(b) shows the temperature dependence of total relaxation time $t_{\text{total}}$ and $\Delta_m$, which disappear above 80 K. We clarify that the measured thermodynamic response is influenced by how quickly the sample is heated up, which is dependent on the specific heat capacity of the sample holder at different temperatures. Hence the relaxation time at lower temperatures is underestimated; however, it does not affect the qualitative conclusion that cAFM wires have collective domain wall behaviors. We also note no such anomalous response is observed in the 2D Hall bar device at any temperature (Fig. 4b).

In summary, we have investigated the thermodynamic responses of hexagonal and straight cAFM nanowire devices. As opposed to the 2D Hall bar device which shows regular ohmic behaviors, resistance ratios in the hexagonal device diverge from cool-down to cool-down, with the observation of characteristic temperatures matching those of ferroelastic domain walls in STO. Statistics on the resistance ratios reveal these behaviors are unlikely from interacting with insulating *Z*/*X*,*Y* domain walls in STO substrate; instead, they originate intrinsically from the cAFM nanowires. Further study on a straight wire device shows the cAFM nanowire experiences an anomalous, slowly recovering resistance change upon thermal stimulus. These evidence point to the domain wall nature of cAFM nanowires, which is crucial to understand the novel physics, including electron pairing and high mobility transport on cAFM nanowires. The microscopic origin of the electric and thermodynamic responses is not clear, but should generally agree with domain wall movement or trapping and de-trapping the oxygen vacancies on domain walls. Provided that abnormal thermodynamic response is absent in the 2D Hall bar device, it is plausible to conclude that ferroelastic distortion under cAFM wires is significant as opposed to 2D interface. An ultimate study would involve real-space imaging of cAFM nanowires at cryogenic temperatures, e.g. PFM, to directly relate the morphology of cAFM nanowires to quantum transport.

**Acknowledgement**

This work is supported by National Natural Science Foundation of China under grant No. NSFC 11874054 and the University of Science and Technology of China under grant No. WK3540000003, KY2030000160 and YD3540002001.

**APPENDIX A: SAMPLE GROWTH AND CAFM DEVICE FABRICATION**

LAO/STO samples are grown by pulsed laser deposition on (001) STO substrates. Low mis-cut angle (<0.1°) single-crystal SrTiO$_3$ substrate is TiO$_2$-terminated by etching in buffered HF for 30 s. Atomically smooth surface is formed by annealing at 960 °C for 90 min. A 3-unit-cell LaAlO$_3$ film is grown on the surface of SrTiO$_3$ by PLD with 1 Hz pulse frequency at a temperature of 710 °C and 5×10$^{-5}$ mbar oxygen pressure, while layer by layer growth is monitored by the reflection high-energy electron diffraction (RHEED) during growth. Then the film is gradually cooled down to room temperature. The sample is patterned with interface electrodes by Ar+ ion etching 25 nm and backfilling with Au/Ti (20 nm/5 nm) contacting the LaAlO$_3$/SrTiO$_3$ interface.

The sample canvas which is an area typically 25 μm ×25 μm defined by interface electrodes is erased first by a cAFM tip with $V_{\text{tip}} = -10$ V at 2 μm/s speed densely prior to device writing. Then funnel shape virtual electrodes are written at 2 μm/s speed with $V_{\text{tip}} = 15$ V to interface the metal electrodes. Wire leads contacting the virtual electrodes are then drawn to the device area at 400 nm/s speed and $V_{\text{tip}} = 15$ V. Finally, the hexagonal or straight main channel is written at a slower writing speed of 300 nm/s with the same voltage. After the device is finished, the sample is quickly transferred to cryostat for measurements. The sample, heater, and thermometer are mounted closely together on the same cold finger made of oxygen-free copper, so that temperature differences and lags in reading are negligible.

Over 10 devices have been fabricated, with 4 presented in this work: device A and B in the main text, and device C and D in Fig. 5 to demonstrate random resistance jerks.

Device C has the same geometry as device A in the main text, and device D is a straight wire but orientated perpendicular to device B.

**APPENDIX B: ADDITIONAL DEVICES WITH RESISTANCE JERKS**

It is not rare that resistance jerks show up in some devices even with similar geometries with Device A and B below $T_{c2}$=80 K. Figure 5 shows resistance ratios between sections of wires in device C and D. It is obvious that the ratios are bundling together above $T_{c2}$ and spreading out with random resistance jerks at lower temperatures (Fig. 5(b) and Fig. 5(d)). Such phenomena can be understood in the framework of this work that when domain wall gets increasingly mobile[27], energy jerks resulting from domain wall pinning and depinning events in the non-equilibrium thermodynamics causes the observed random resistance switching[24,36].

**APPENDIX C: DATA FOR ADDITIONAL RESISTANCE RATIOS IN HEXAGONAL DEVICE**

The rest of the resistance ratios data and the statistics counting (Fig.6) in the hexagonal wire device. In most of the resistance ratios data, such behavior which shows spreading out below 80 K can be observed. Almost all the curves are performing as single-peak shapes which is consistent with the main text description, though only two of ten counting curves show significant fluctuation.

From another perspective, it is shown by Goble *et al*. that the transport measurement in 2D channel along or across domain walls causes observable resistance change[29]. However, in our experiment, the resistances of wire sections in different orientations in the hexagonal device are almost the same. This also suggests that domain wall characteristics of sketched cAFM nanowires are less likely originated from intersecting with the bulk domain walls.

**APPENDIX D: RESISTANCE DIFFERENCES ANALYSIS IN HEXAGONAL DEVICE A**

In the data analysis, we also tried to analyze the data with resistance differences (Fig. 7) to remove the common contribution of resistance, which yielded qualitatively the same results. Namely, normalized resistance differences stay the same above $T$=80 K in all the cool-downs, but then spread out abnormally as a result of domain redistribution.

**APPENDIX E: DEVICE SCHEMATIC OF DEVICE B**

The device consists of three sections of 4 μm length wires with the same lithography parameter (Fig. 8(a)). All three sections of wires show the same thermodynamic response of resistance anomalies with the first section of wire data presented in this work.

**APPENDIX F: DETERMINATION OF THERMODYNAMIC DURATION IN DEVICE B**

The thermodynamic response of resistance anomalies appears below 80 K and the exponential relaxation process is more pronounced at lower temperatures. The anomalous resistance drop and heating effect compete. We define the ending of this anomalous response by looking at the zero crossing of the second derivative $d^2R/dt^2$ curve, where the anomalous resistance drop and regular heating effect balance. This will underestimate the total response time $t_{total}$; however, it is good enough to draw the conclusions in this work. Figure 9 shows how the blue dashed line is extracted in the main text according to the method described above. In Fig. 9(b), no ending point in $d^2R/dt^2$ could be identified at $T$=90 K (or above), which marks the absence of anomalous thermodynamic response at this temperature.

**APPENDIX G: THERMODYNAMIC RESPONSE IN 2D LAO/STO HALL BAR**

We repeated the two experiments in the 2D LAO/STO Hall bar device in dimension of $25\ \mu m \times 20\ \mu m$ (Fig. 10 (a)). The 5-unit-cell LAO/STO sample is grown by PLD under the same parameters which are described in the sample growth section (APPENSIX A). The Hall bar device is fabricated through standard optical lithography, and transport measurement is performed in the same setup of the nanowire measurement.

The resistance ratios of different temperature cycles in the same hall bar stay closely in all temperature ranges rather than the spreading behavior (Fig. 10 (b)), where the $r^*(T) = r(T)/r(145\ K)$ and $r(T) = R(T)/R_{1^{st}\ \text{cool-down}}(T)$. Moreover, the resistance of the Hall bar displays a normal ohmic behavior (Fig. 10 (c) and (d)) rather than the anomalous resistance drop in the cAFM nanowires. This is understood through this work that the ferroelastic distortion under the 2D interface is negligible, while the cAFM nanowires are subject to ferroelastic distortion along the whole wire.

**Figures:**

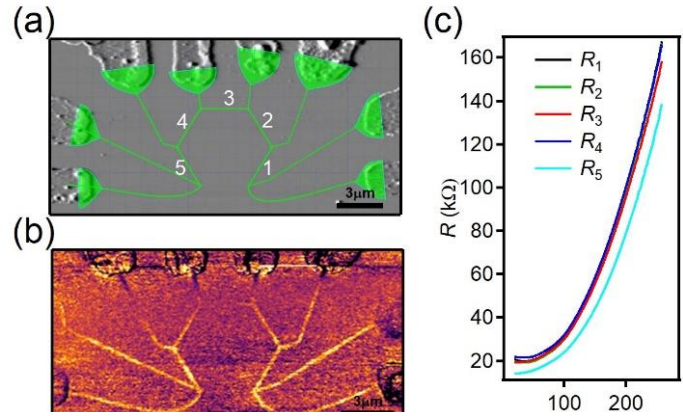

Fig.1 Hexagonal wire device. (a) Device schematic overlaid on an AFM canvas image. Green color denotes wires and electrodes to write. Wire sections are labelled as 1, 2, 3, 4 and 5. (b) PFM imaging reveals the written structures. (c) Temperature dependence of $R_1$, $R_2$, $R_3$, $R_4$ and $R_5$.

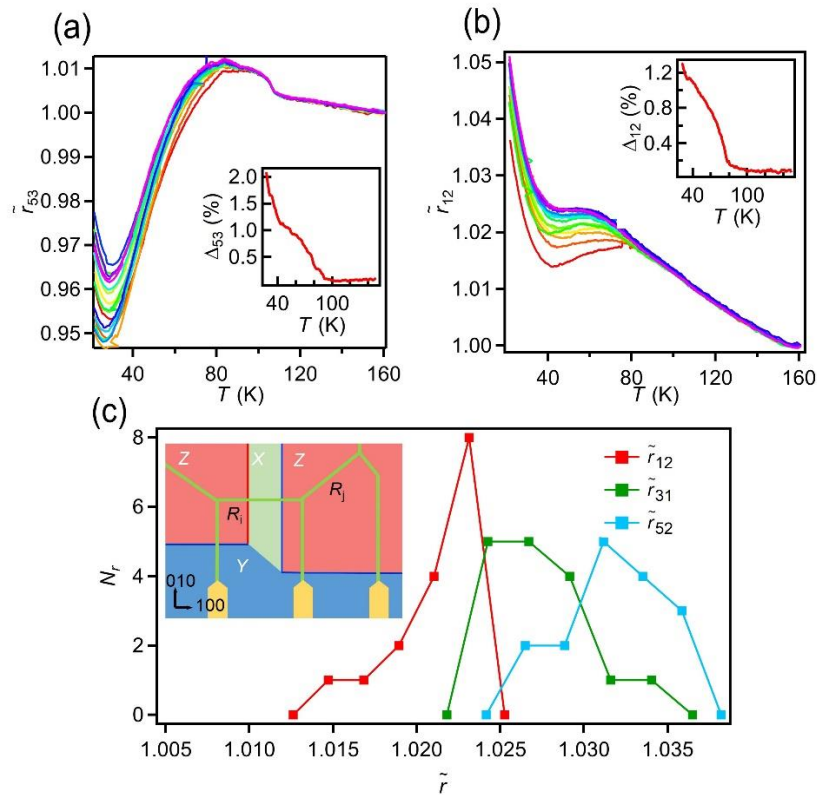

Fig. 2. Resistance anomaly in different temperature cycles. (a) and (b) Normalized resistance ratios of $R_5/R_3$ and $R_1/R_2$ as a function of temperature, respectively. The insets show the spreading between maximum and minimum ratio values $\tilde{r}_{mn}(T)$. (c) Occurrence frequencies of $\tilde{r}_{31}$, $\tilde{r}_{15}$ and $\tilde{r}_{52}$ at 50 K. The inset shows a configuration of (2,0) in which one wire $R_i$ intersects with domain walls twice while the other wire $R_j$ does not intersect with a domain wall.

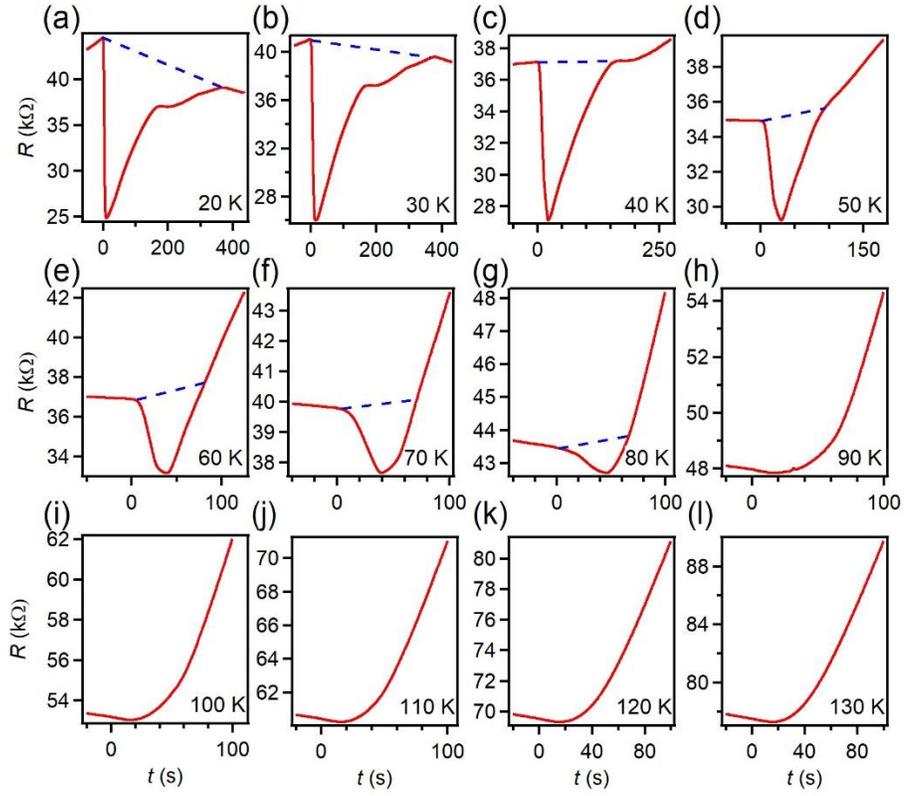

Fig. 3 Thermodynamical response of the sketched cAFM nanowire in device B. (a) - (l) The time evolution of resistances at 20 K to 130 K in step of 10 K. The blue dashed guide lines qualitatively follow the normal warm up resistance response.

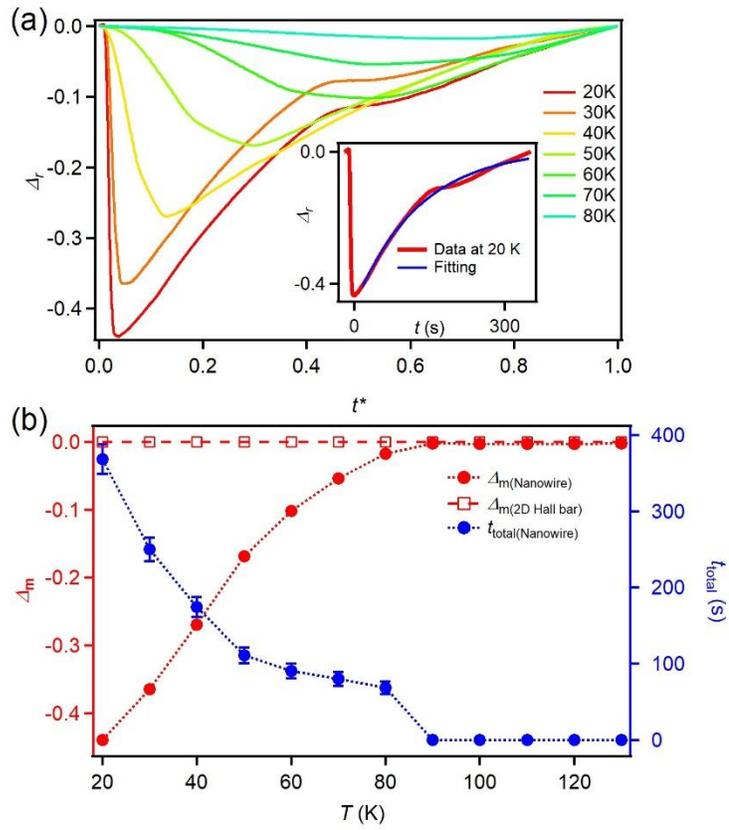

Fig. 4 Anomalous resistance drop and relaxation time in device B. (a) The anomalous resistance drop $\Delta_r$ decreases with increasing temperature. Here the $t^*$ axis is normalized by total relaxation time. (b) Maximum resistance changes and the total relaxation time as a function of temperature for the nanowire and Hall bar devices.

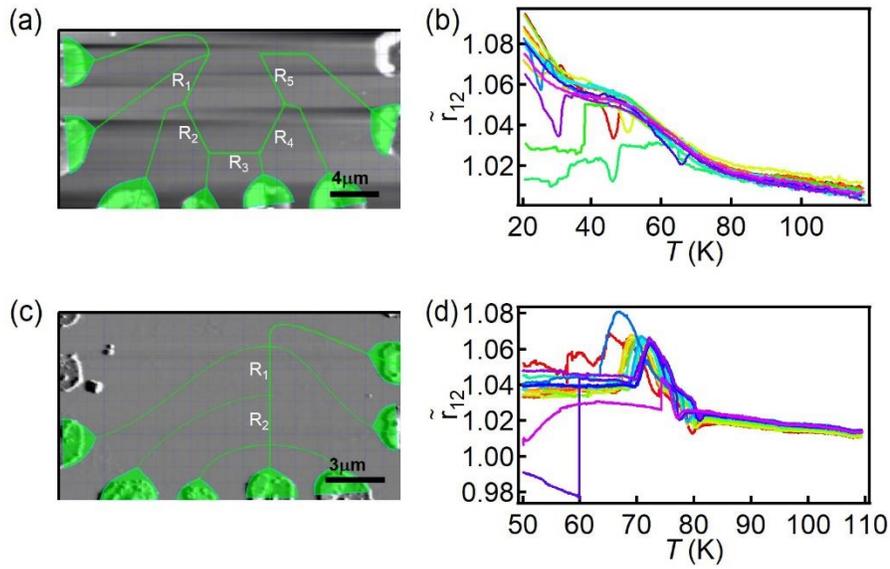

Fig. 5 Resistance jerks in device C and D. (a) and (c) show devices geometries of device C and D. (b) and (d) The corresponding resistance in 13 and 21 temperature cycles respectively, which shows similar behavior as device A, but with additional resistance jerks.

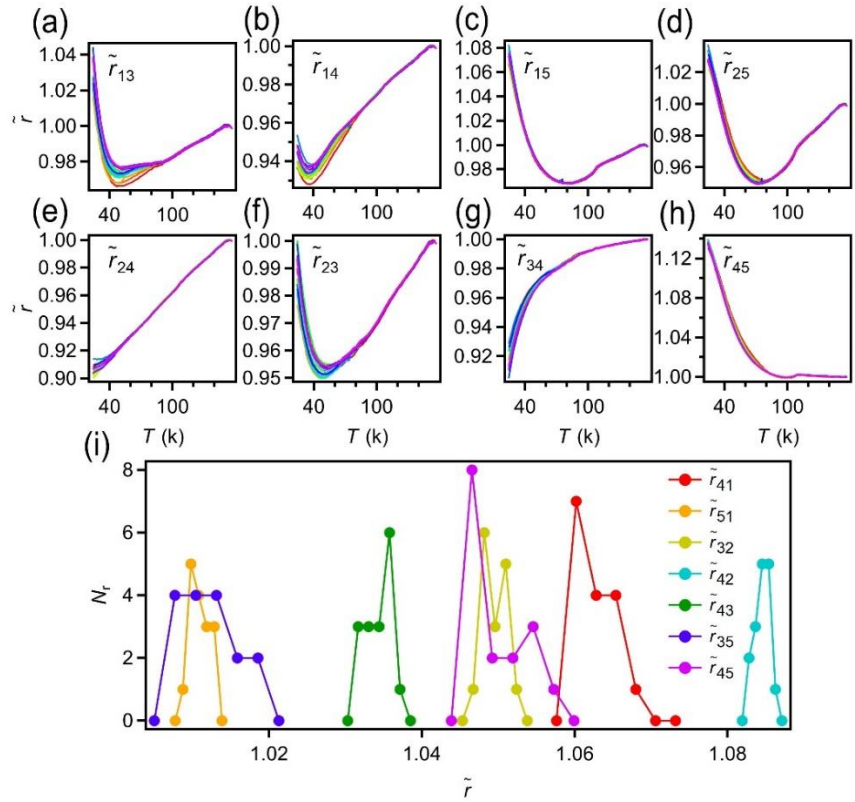

Fig. 6 (a~h) Additional resistance ratio data for 17 times temperature cycles in device A. (i) Rest of occurrence frequencies in hexagonal device A.

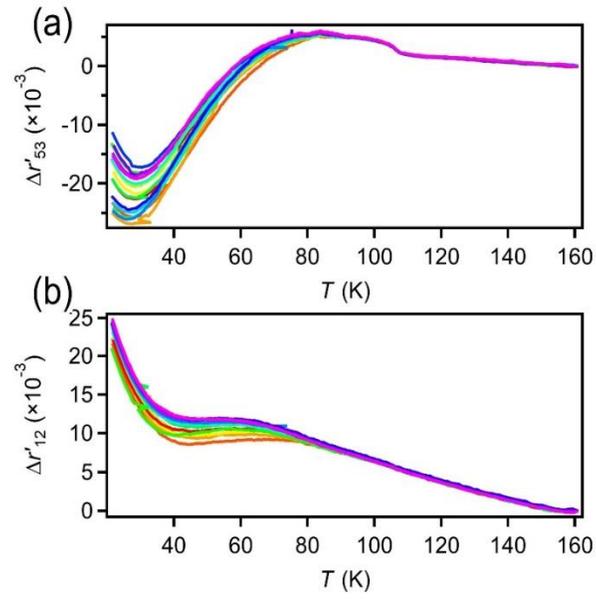

Fig. 7 Resistance spreading analysis in different temperature cycles by resistance difference. (a) and (b) show the same spreading behavior in resistance differences below 80 K as Fig. 2 in the main text, where $\Delta r'_{ij}(T) = \Delta r_{ij}(T)/\Delta r_{ij}(160\ \text{K})$ and $\Delta r_{ij} = \frac{R_i - R_j}{R_i + R_j}$.

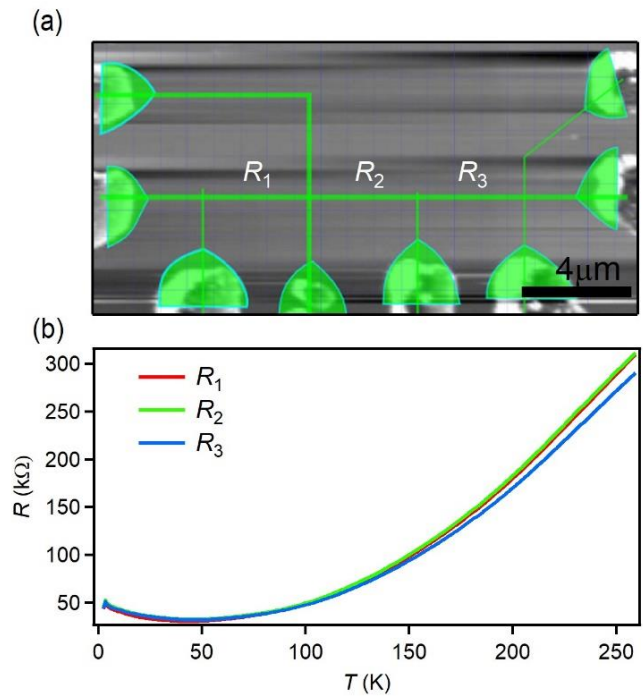

Fig. 8. Straight wire device B and transport characterization. (a) Device schematic showing three sections of wires with 4 μm long each. (b) Temperature dependence of resistance in each section of wires.

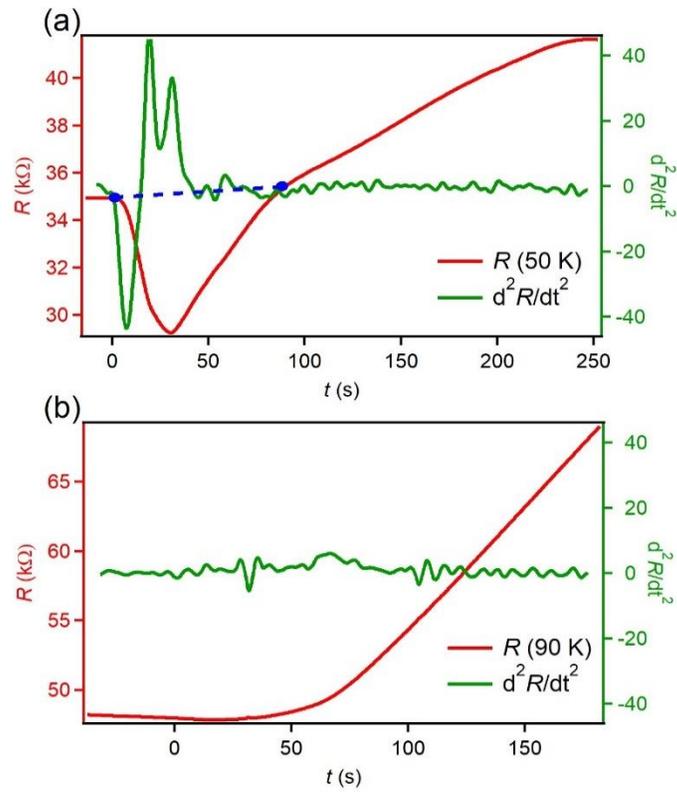

Fig. 9 Definition of total thermodynamic response time. (a) and (b) show the data at 50 K and 90 K respectively. The red curves are the resistance as a function of time and the green curves are the second-order derivative of resistance versus time. The blue dashed line indicates the whole process from heating up to relaxation, corresponding to the blue dashed line in Fig. 3.

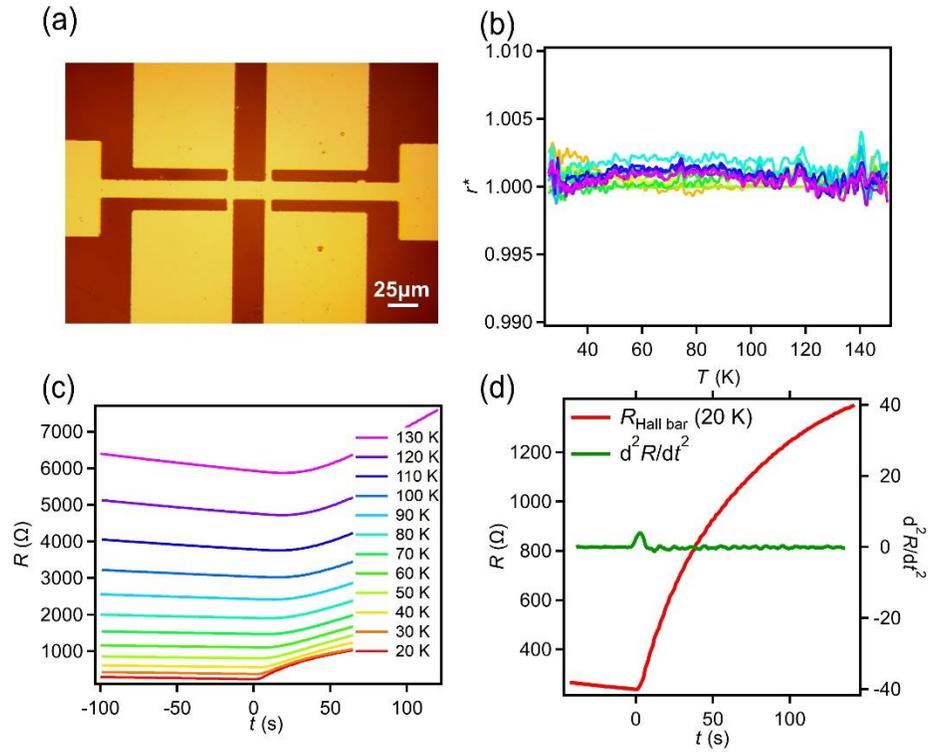

Fig. 10. Thermodynamic response and temperature dependent resistance in a 2D LAO/STO Hall bar device. (a) Optical image of 2D LAO/STO Hall bar device. (b) shows the resistance ratios of different temperature cycles in the same Hall bar. No abnormal resistance spreading is observed. (c) shows the time evolution of resistance after the thermal stimulus at $t = 0$ from 20 K to 130 K. No resistance drop is observed. (d) shows the definition of total thermodynamic response time at 20K. The method is the same with that in the Fig. 9(b).

**References:**

[1] A. Ohtomo and H. Y. Hwang, Nature **427**, 423 (2004).
[2] N. Reyren, S. Thiel, A. D. Caviglia, L. F. Kourkoutis, G. Hammerl, C. Richter, C. W. Schneider, T. Kopp, A. S. Ruetschi, D. Jaccard *et al.*, Science **317**, 1196 (2007).
[3] S. Thiel, G. Hammerl, A. Schmehl, C. W. Schneider, and J. Mannhart, Science **313**, 1942 (2006).
[4] M. Ben Shalom, M. Sachs, D. Rakhmilevitch, A. Palevski, and Y. Dagan, Phys. Rev. Lett. **104**, 126802 (2010).
[5] A. D. Caviglia, M. Gabay, S. Gariglio, N. Reyren, C. Cancellieri, and J. M. Triscone, Phys. Rev. Lett. **104**, 126803 (2010).
[6] A. Brinkman, M. Huijben, M. Van Zalk, J. Huijben, U. Zeitler, J. C. Maan, W. G. Van der Wiel, G. Rijnders, D. H. A. Blank, and H. Hilgenkamp, Nat. Mater. **6**, 493 (2007).
[7] C. Cen, S. Thiel, G. Hammerl, C. W. Schneider, K. E. Andersen, C. S. Hellberg, J. Mannhart, and J. Levy, Nat. Mater. **7**, 298 (2008).
[8] C. Cen, S. Thiel, J. Mannhart, and J. Levy, Science **323**, 1026 (2009).
[9] G. L. Cheng, J. P. Veazey, P. Irvin, C. Cen, D. F. Bogorin, F. Bi, M. C. Huang, S. C. Lu, C. W. Bark, S. Ryu *et al.*, Phys. Rev. X **3**, 011021 (2013).
[10] J. P. Veazey, G. L. Cheng, P. Irvin, C. Cen, D. F. Bogorin, F. Bi, M. C. Huang, C. W. Bark, S. Ryu, K. H. Cho *et al.*, Nanotechnology **24**, 375201 (2013).
[11] G. L. Cheng, P. F. Siles, F. Bi, C. Cen, D. F. Bogorin, C. W. Bark, C. M. Folkman, J. W. Park, C. B. Eom, G. Medeiros-Ribeiro *et al.*, Nat. Nanotechnol. **6**, 343 (2011).
[12] G. L. Cheng, M. Tomczyk, S. C. Lu, J. P. Veazey, M. C. Huang, P. Irvin, S. Ryu, H. Lee, C. B. Eom, C. S. Hellberg *et al.*, Nature **521**, 196 (2015).
[13] A. Annadi, G. L. Cheng, H. Lee, J. W. Lee, S. C. Lu, A. Tylan-Tyler, M. Briggeman, M. Tomczyk, M. C. Huang, D. Pekker *et al.*, Nano Lett. **18**, 4473 (2018).
[14] G. L. Cheng, A. Annadi, S. C. Lu, H. Lee, J. W. Lee, M. C. Huang, C. B. Eom, P. Irvin, and J. Levy, Phys. Rev. Lett. **120**, 076801 (2018).
[15] M. Tomczyk, G. L. Cheng, H. Lee, S. C. Lu, A. Annadi, J. P. Veazey, M. C. Huang, P. Irvin, S. Ryu, C. B. Eom *et al.*, Phys. Rev. Lett. **117**, 096801 (2016).
[16] L. Fidkowski, H. C. Jiang, R. M. Lutchyn, and C. Nayak, Phys. Rev. B **87**, 014436 (2013).
[17] P. Irvin, J. P. Veazey, G. L. Cheng, S. C. Lu, C. W. Bark, S. Ryu, C. B. Eom, and J. Levy, Nano Lett. **13**, 364 (2013).
[18] M. Briggeman, M. Tomczyk, B. B. Tian, H. Lee, J. W. Lee, Y. C. He, A. Tylan-Tyler, M. C. Huang, C. B. Eom, D. Pekker *et al.*, Science **367**, 769 (2020).
[19] Y. Y. Pai, H. Lee, J. W. Lee, A. Annadi, G. L. Cheng, S. C. Lu, M. Tomczyk, M. C. Huang, C. B. Eom, P. Irvin *et al.*, Phys. Rev. Lett. **120**, 147001 (2018).
[20] F. Bi, D. F. Bogorin, C. Cen, C. W. Bark, J. W. Park, C. B. Eom, and J. Levy, Appl. Phys. Lett. **97**, 173110 (2010).
[21] M. C. Huang, F. Bi, S. Ryu, C. B. Eom, P. Irvin, and J. Levy, APL Mater. **1**, 052110 (2013).
[22] P. A. Fleury, J. F. Scott, and J. M. Worlock, Phys. Rev. Lett. **21**, 16 (1968).
[23] K. A. Muller, W. Berlinger, and F. Waldner, Phys. Rev. Lett. **21**, 814 (1968).
[24] J. F. Scott, E. K. H. Salje, and M. A. Carpenter, Phys. Rev. Lett. **109**, 187601 (2012).
[25] E. K. H. Salje, O. Aktas, M. A. Carpenter, V. V. Laguta, and J. F. Scott, Phys. Rev. Lett. **111**, 247603 (2013).
[26] M. Honig, J. A. Sulpizio, J. Drori, A. Joshua, E. Zeldov, and S. Ilani, Nat. Mater. **12**, 1112 (2013).
[27] Y. Frenkel, N. Haham, Y. Shperber, C. Bell, Y. W. Xie, Z. Y. Chen, Y. Hikita, H. Y. Hwang, E. K. H. Salje, and B. Kalisky, Nat. Mater. **16**, 1203 (2017).

[28]	B. Kalisky, E. M. Spanton, H. Noad, J. R. Kirtley, K. C. Nowack, C. Bell, H. K. Sato, M. Hosoda, Y. W. Xie, Y. Hikita *et al.*, Nat. Mater. **12**, 1091 (2013).
[29]	N. J. Goble, R. Akrobetu, H. Zaid, S. Sucharitakul, M. H. Berger, A. Sehirlioglu, and X. P. A. Gao, Sci. Rep. **7**, 44361 (2017).
[30]	D. Pekker, C. S. Hellberg, and J. Levy, (2020), p. arXiv:2002.11744.
[31]	H. J. H. Ma, S. Scharinger, S. W. Zeng, D. Kohlberger, M. Lange, A. Stohr, X. R. Wang, T. Venkatesan, R. Kleiner, J. F. Scott *et al.*, Phys. Rev. Lett. **116**, 257601 (2016).
[32]	E. K. H. Salje, A. Buckley, G. Van Tendeloo, Y. Ishibashi, and G. L. Nord, Am. Mineral. **83**, 811 (1998).
[33]	R. J. Harrison and E. K. H. Salje, Appl. Phys. Lett. **97**, 021907 (2010).
[34]	S. Puchberger, V. Soprunyuk, W. Schranz, A. Troster, K. Roleder, A. Majchrowski, M. A. Carpenter, and E. K. H. Salje, APL Mater. **5**, 046102 (2017).
[35]	B. Casals, G. F. Nataf, D. Pesquera, and E. K. H. Salje, APL Mater. **8**, 011105 (2020).
[36]	D. Pesquera, M. A. Carpenter, and E. K. H. Salje, Phys. Rev. Lett. **121**, 235701 (2018).
[37]	S. K. Ojha, S. Hazra, P. Mandal, R. K. Patel, S. Nigam, S. Kumar, and S. Middey, Phys. Rev. Appl. **15**, 054008 (2021).